\documentclass[review]{elsarticle}
\usepackage{lineno,hyperref}
\usepackage[english]{babel}
\usepackage[numbers]{natbib}
\usepackage{xcolor}
\usepackage{latexsym,amsmath,amssymb,amsbsy,graphicx,geometry}
\usepackage{epstopdf}
\usepackage{epsfig}
\modulolinenumbers[5]

\begin{document}
	
\begin{frontmatter}

\title{Mechanical Response of Mesoporous Amorphous NiTi Alloy to External Deformations}

\author[adr1,adr2]{Bulat N. Galimzyanov}
\ead{bulatgnmail@gmail.com}
\author[adr1,adr2]{Anatolii V. Mokshin}
\ead{anatolii.mokshin@mail.ru}

\address[adr1]{Kazan Federal University, 420008 Kazan, Russia}
\address[adr2]{Udmurt Federal Research Center of the Ural Branch of the Russian Academy of Sciences, 426067 Izhevsk, Russia}

\begin{abstract}
The porous titanium nickelide is very popular in various industries due to unique combination of physical and mechanical properties such as shape memory effect, high corrosion resistance, and biocompatibility. The non-equilibrium molecular dynamics simulation was applied to study the influence of porosity degree on mechanical properties of porous amorphous titanium nickelide at uniaxial tension, uniaxial compression, and uniform shear. We have found that the porous amorphous alloy is characterized by a relatively large value of Young's modulus in comparison to its crystalline analogue. It has been found that the system with a percolated network of pores exhibits improved elastic characteristics associated with resistance to tensile and shear. The system contained isolated spherical pores is more resistant to compression and less resistant to tensile and shear. These results can be applied to develop and improve the methods for making amorphous metal foams.
\end{abstract}


\end{frontmatter}

\section{Introduction}

Amorphous metal foams based on Ni and Ti exhibit unique physical and mechanical properties, among which are low specific weight, low stiffness, high corrosion resistance and good biocompatibility~\cite{Lu_Xiao_1999,Zhao_2012,Qiu_Young_2015,Garcia_Moreno_2016,Beltyukov_Ladyanov_2019,Muhammad_Virginia_2001}. These properties depend primarily on size and geometry of the pores as well as on the stability of an amorphous matrix~\cite{Simoneau_2016,Atwater_Guevara_2018,Nguyen_Jang_2019,Rammerstorfer_Pahr_2006}.
Existing methods for production of the porous amorphous metallic alloys, such as the spark plasma sintering and three-stage synthesis method, usually make it possible to obtain the materials with the pores greater than $50$~nm~\cite{Schroers_Veazey,Schroers_Veazey_2003}. These methods do not guarantee the complete absence of crystalline nuclei, which can start to grow and destroy the amorphous structure of porous material. The cooling procedure of an equilibrium liquid melt with the ultrafast cooling rates (higher than $10^{6}$~K/s) and ultrafast heat dissipation can be applied to obtain a porous material with the  amorphous structure and to prevent the crystal nucleation events~\cite{Galimzyanov_Mokshin_JCG_2019,Mokshin_Khusnutdinoff_PCCP_2020,Galimzyanov_Mokshin_AM_2019}. In practice, these conditions are difficult to be realized for the most known metallic alloys~\cite{Sanditov_Badmaev_2019,Yu_Eifert_1998,Betts_2012,Tantavisut_2018,Contuzzi_2019}.

Titanium nickelide alloy (Ni$_{50}$Ti$_{50}$ or NiTi) is the most famous functional material among intermetallic compounds especially due to the shape memory effect of this alloy~\cite{Elahinia_Hashemi_2012,Shelyakov_2013,Sepe_Auricchio_2016,Kalashnikov_Koledov_2018,Sellitto_Riccio_2019,Gardan_2019,Tsouknidas_2019}. Laser and spark plasma sintering are the main methods of powder metallurgy for synthesis of crystalline NiTi with micro- and macroporous structures~\cite{Shishkovsky_Yadroitsev_2012,Ye_Liu_1998,Dudina_Bokhonov_2019}. The relative simplicity and accessibility of these methods make it possible to study the mechanical properties of this alloy. However, it is difficult to obtain the porous amorphous NiTi using the above methods primarily due to the high melting point of titanium nickelide, $T_{m}\simeq1600$~K~\cite{Nitinol_properties,Anikeev_Hodorenko_2017}. Absence of comprehensive experimental measurements and numerical simulations is the reason why the physical and mechanical properties of porous amorphous NiTi are still poorly studied~\cite{Lefebvre_Dunand_2008,Rivard_Prokoshkin_2014,Sawei_Xinna_2015,Smolin_Makarov_2016,Deng_Wang_2019}. Improved computational methods and models are required to reproduce the structure and properties of the porous amorphous NiTi~\cite{Sepe_Auricchio_2016,Panico_Brinson_2008}.

In the present work, we construct a model of porous amorphous NiTi through rapid cooling of a low-density liquid melt. The effect of uniaxial tension, uniaxial compression and uniform shear on the porous amorphous samples with different porosity is studied. We show that the mechanical characteristics of porous amorphous NiTi differ significantly from the mechanical properties of its crystalline analogue. This is confirmed by comparing our simulation results with the available experimental data.

\section{Preparation of porous amorphous NiTi}

The ground crystalline state of NiTi at the temperature $T=0$~K is characterized by the B2 type cubic lattice. The Ni and Ti atoms are located inside the simulation box with the same length of edges $L_{x}=L_{y}=L_{z}\simeq12.4$~nm. The periodic boundary conditions are applied in all the directions. The system consists of $N=109\,744$ atoms: $54\,872$ atoms of Ni and the same number of Ti atoms. Interaction energies between the atoms are determined by the modified embedded-atom me\-thod (MEAM) potential developed by Ko et al. for NiTi~\cite{Ko_2015}. As found before in Ref.~\cite{Galimzyanov_Mokshin_FTT_2020}, this potential correctly reproduces the structural and dynamic properties of NiTi.

According to the MEAM formulation~\cite{Lee_Baskes_2000}], the total energy $E$ of the considered many-particle system is determined as
\begin{equation}\label{eq_meam_1}
E=\sum_{i=1}\left(F_{i}(\bar{\rho}_{i}(r))+\frac{1}{2}\sum_{j\neq i}^{N}\phi_{ij}(r)\right).
\end{equation}
Here, $\phi_{ij}(r)$ is the pair interaction between atoms $i$ and $j$ separated by a distance $r$. The embedding function $F_{i}(\bar{\rho}_{i}(r))$ is defined in the form
\begin{equation}\label{eq_meam_2}
F_{i}(\bar{\rho}_{i}(r))=AE_{s}\bar{\rho}_{i}(r)\ln\bar{\rho}_{i}(r).
\end{equation}
The parameter $A$ is the scaling factor for the embedding function and $E_S$ is the sublimation energy. The effective electron density $\bar{\rho}_{i}(r)$ is given by
\begin{equation}\label{eq_meam_3}
\bar{\rho}_{i}(r)=2\rho_{i}^{(0)}(r)\left[1+\exp\left(-\sum_{l=1}^{3}t^{(l)}\left[\frac{\rho_{i}^{(l)}(r)}{\rho_{i}^{(0)}(r)}\right]^{2}\right)\right]^{-1},
\end{equation}
where $\rho_{i}^{(0)}(r)$ is the spherically symmetric partial electron density; $\rho_{i}^{(l)}(r)$ ($l=1,\,2,\,3$) are the angular contributions (see Ref.~\cite{Lee_Baskes_2000}). According to Ref.~\cite{Lee_Baskes_2000}, the quantities $\rho_{i}^{(0)}(r)$ and $\rho_{i}^{(l)}(r)$ depend on the atomic electron density $\rho_{j}^{a(l)}(r)$ of $j$th atom at a distance $r$ from $i$th atom:
\begin{equation}\label{eq_meam_4}
\rho_{j}^{a(l)}(r)=\exp\left(-\beta^{(l)}\left[\frac{r}{r_{e}}-1\right]\right).
\end{equation}
Here, $t^{(l)}$ and $\beta^{(l)}$ represent the weight factors and the dumping coefficients for the partial electron densities; $r_e$ is the equilibrium nearest-neighbor distance in an equilibrium reference structure. In the MEAM, only the first nearest-neighbor interactions are considered. Therefore, the pair potential $\phi(r)$ between two atoms separated by a distance $r$ is defined in the form
\begin{equation}\label{eq_meam_5}
\phi(r)=\frac{2}{Z}\left[E^{u}(r)-F(\bar{\rho}_{0}(r))\right],
\end{equation}
where
\begin{equation}\label{eq_meam_6}
E^{u}(r)=-E_{S}(1+\alpha+d\alpha^{3})e^{-\alpha},  
\end{equation}
\begin{equation}\label{eq_meam_7}
\alpha=\left(\frac{9BV}{E_{S}}\right)^{1/2}\left[\frac{r}{r_e}-1\right].
\end{equation}
Here, $E^{u}(r)$ is the universal function for a uniform expansion or contraction in the reference structure~\cite{Rose_1984}; $Z$ is the number of nearest-neighbor atoms; $\bar{\rho}_{0}$ is the background electron density for the reference structure; $B$ is the bulk modulus; $V$ is the atomic volume of solid elements; $d$ is an adjustable parameter. The values of the quantities $t^{(l)}$, $\beta^{(l)}$, $r_e$, $E_S$, $A$, $B$, $V$ and $d$ are given in Ref.~\cite{Ko_2015} for pure Ni and Ti systems as well as for NiTi alloy.

Initially, the crystalline sample has been melted at the constant pressure $p=1$~atm by heating above the melting (liquidus) temperature $T_{m}\simeq1600$~K. The equilibrated liquid samples at the temperatures $T=3500$, $6500$, $7000$, $7500$, $8000$ and $8600$~K have been prepared. Note that all these temperatures do not exceed the boiling temperature $T_{b}\simeq8750$~K of NiTi~\cite{Nikiforov_2019}.

To generate the homogeneous amorphous sample without pores, a liquid melt at the temperature $T=3500$~K has been isobarically cooled with the rate $10^{13}$~K/s to a state with the temperature $T=300$~K at the pressure $p=1$~atm. The generated amorphous sample has the density $\rho_{0}\simeq6.17$~g/cm$^{3}$  that is close to the density of crystalline NiTi, $\simeq6.45$~g/cm$^{3}$, at the same temperature~\cite{Duerig_Pelton_1994}. Note that the glass transition temperature $T_g$ of titanium nickelide estimated by T.~Rouxel and Y.~Yokoyama is $\sim 750$~K~\cite{Rouxel_Yokoyama_2015}.

\begin{figure}[tbh]
	\centering
	\includegraphics[width=0.65\linewidth]{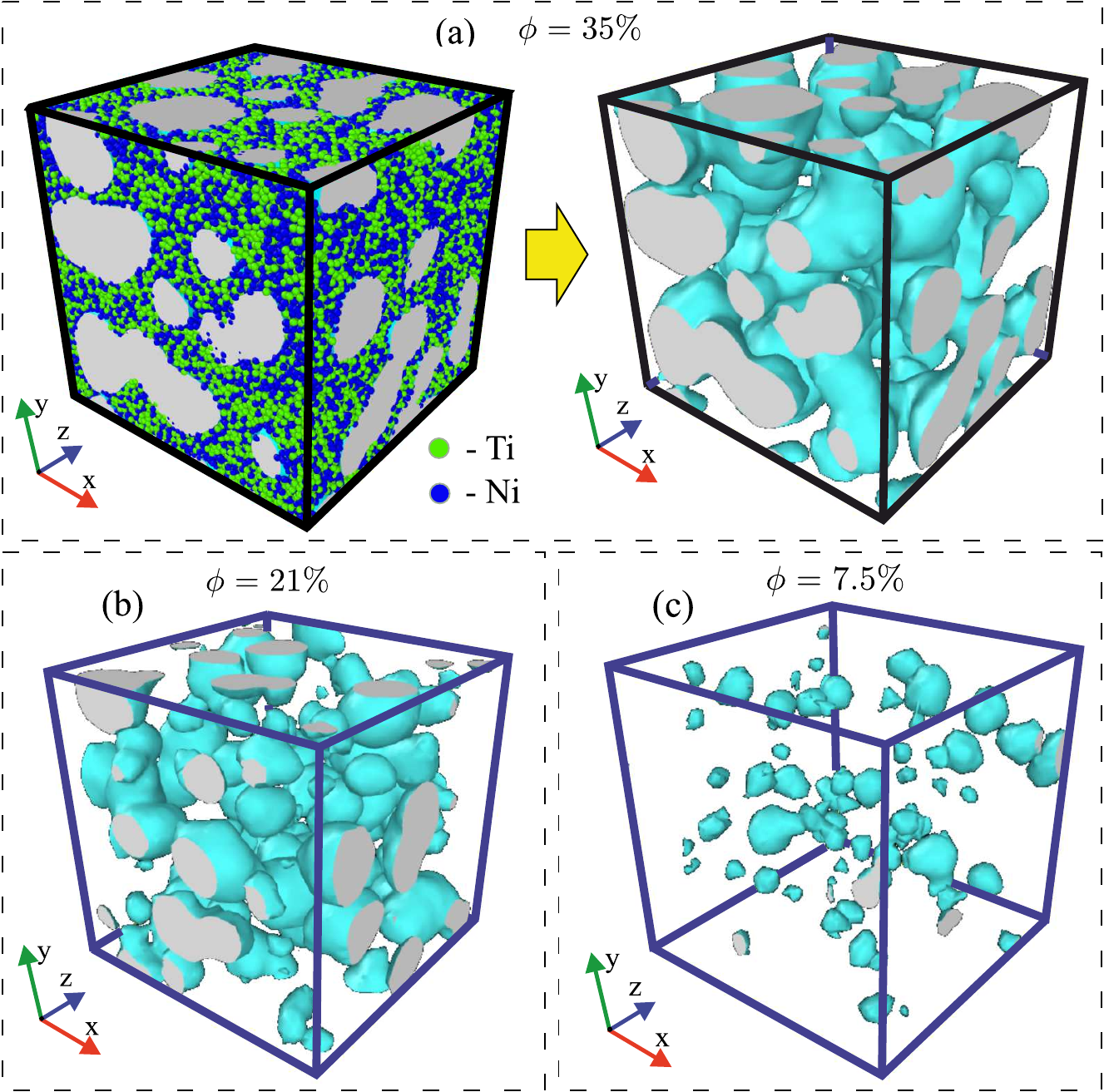}
	\caption{(color online) Snapshots of amorphous NiTi at the different degrees of porosity: (a) $\phi=35$\%, (b) $\phi=21$\% and (c) $\phi=7.5$\%. On the left side of panel (a), it is shown a simulation porous sample. On the right side of panel (a) as well as in panels (b) and (c), only the three-dimensional images of the pore walls are shown. The 3D images of pores are constructed by using the Stukowski's computational algorithm~\cite{Stukowski_2014}.}
	\label{fig_1}
\end{figure}

Amorphous samples with various porosity have been prepared by isochoric cooling of melt samples equilibrated at the temperatures $T=6500$, $7000$, $7500$, $8000$ and $8600$~K. Each sample has been cooled with the rate $10^{13}$~K/s to the temperature $300$~K. It should be noted that with such ultrafast cooling, a negative internal pressure arises in the melt, and the solidifying samples do not form a dense homogeneous solid phase. As a result, the melt transforms into porous amorphous solid, where the linear size of pores varies within the interval from $2$~nm to $10$~nm [Fig.~\ref{fig_1}]. After ultrafast cooling, each  sample has been aged at the state with the temperature $T=300$~K and the pressure $p=1$~atm over the time $1$~ns. Amorphous samples prepared in this way have the different densities $\rho\simeq5.71$, $5.24$, $4.87$, $4.5$ and $4.01$~g/cm$^{3}$. These values of the density $\rho$ are much lower than the density $\rho_{0}$ of the homogeneous amorphous system.

The porosity of the system is estimated by the known ratio~\cite{Dullien_1992}
\begin{equation} \label{eq_porosity}
\phi = \left(1-\frac{\rho}{\rho_{0}}\right)\times100\%.
\end{equation}
In the case of the homogeneous amorphous system, we have $\rho=\rho_{0}$ and the coefficient of porosity is $\phi=0$. The porosity coefficient $\phi$ takes the values $7.5$, $15$, $21$, $27$ and $35$\% for porous amorphous NiTi-samples. Visualization of the porous structure given in Fig.~\ref{fig_1} reveals predominance of the pore networks within the samples at high porosity, $\phi>20$\% [Figs. \ref{fig_1}(a) and~\ref{fig_1}(b)]. The samples with a low degree of porosity (for example, at the porosity $\phi=7.5$\%) contain the isolated pores of a roughly spherical shape [Fig.~\ref{fig_1}(c)].

\section{Amorphous NiTi under various types of deformation}

It is well-known that the internal elastic stresses in the volume of solid arise as a result of external mechanical deformations. These stresses tend to return a deformed solid to its original shape ~\cite{Landau_Lifshitz_1970}. Elastic response is defined by the stress tensor, the nine elements of which are the force components acting on an element of a unit area:
\begin{equation}
\mathbf{\sigma}_{\alpha\beta}(t)\equiv\mathbf{\sigma}_{\alpha\beta}=
\begin{bmatrix}\label{eq_sigma_matrix}
\sigma_{xx} & \sigma_{xy} & \sigma_{xz} \\
\sigma_{yx} & \sigma_{yy} & \sigma_{yz} \\
\sigma_{zx} & \sigma_{zy} & \sigma_{zz}
\end{bmatrix}.
\end{equation}
We evaluate the tensor components $\sigma_{\alpha\beta}$ through the Irwin-Kirkwood equation~\cite{Thompson_2009,Evans_Morriss_2008}
\begin{equation}
\sigma_{\alpha\beta}=-\frac{1}{V}\left(\sum_{i=1}^{N}m_{i}v_{i\alpha}v_{i\beta}+\sum_{i=1}^{N}\sum_{j>i}^{N}r_{ij\alpha}F_{ij\beta}\right).
\end{equation}
Here, the index $\alpha = x,y,z$ characterizes the normal vector of a unit area, and the index $\beta = x,y,z$ corresponds to a force component, $V$ is the volume of a system, $m$ is the mass of the $i$th atom; $v_{i\alpha}$ and $v_{i\beta}$ are the $\alpha$ and $\beta$ components of the peculiar velocity of the $i$th atom; $F_{ij\beta}$ is the $\beta$ component of the force between the particles with the labels $i$ and $j$; $r_{ij\alpha}$ is the
$\alpha$ component of the radius-vector between the particles $i$ and $j$. The diagonal components $\sigma_{xx}$, $\sigma_{yy}$, $\sigma_{zz}$ in Eq.~(\ref{eq_sigma_matrix}) are the normal stresses, which characterize the action of an external force to the unit surface of a solid in the orthogonal direction. The upper and lower symmetric triangular parts of the tensor (\ref{eq_sigma_matrix}) are the shear components that characterize the action of force parallel to the unit surface of solid. 

\begin{figure}[tbh!]
	\centering
	\includegraphics[width=0.5\linewidth]{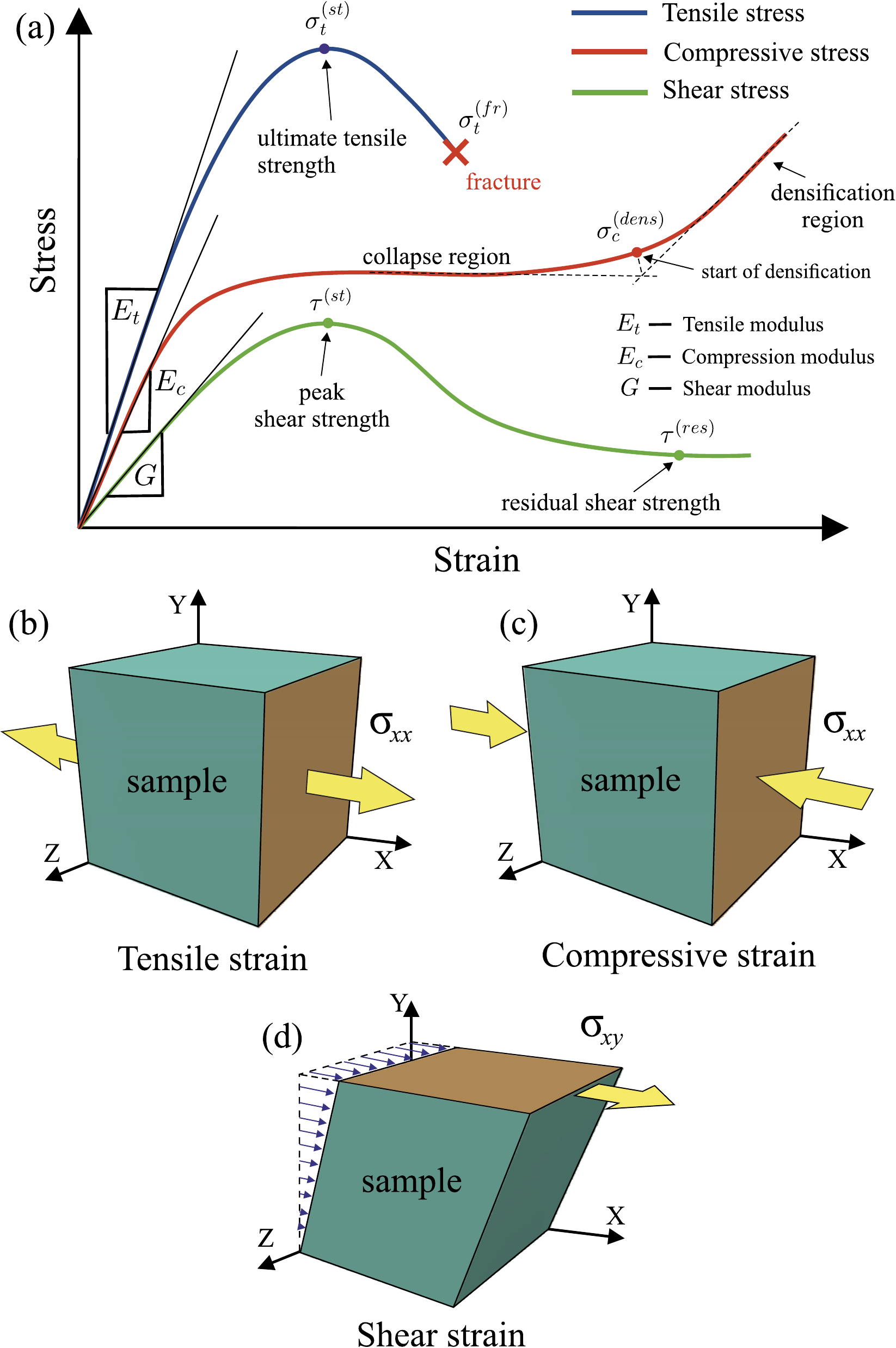}
	\caption{(color online) (a) Typical stress-strain curves for a system under tensile stress, compressive stress and shear stress. Scheme of a system at deformation: (b) by tensile along the $x$-direction, (c) by compression along the $x$-direction, and (d) by shear along the $x$-direction.}
	\label{fig_2}
\end{figure}

Response of considered system to various mechanical deformations is estimated by analysis of the  stress-strain curves. The main elastic moduli such as the tensile modulus (or Young's modulus) $E_{t}$, the compression modulus $E_{c}$ and the shear modulus $G$ are determined by the slope of the linear (elastic) part of the corresponding stress-strain curve [see schematic Fig.~\ref{fig_2}(a)]. The slope of this linear part have weak dependence on the strain rate~\cite{Shen_2012,Tang_2018}. Note that the high elastic moduli will correspond to materials with more pronounced elasticity.

In the given study, the amorphous NiTi samples are exposed to the following types of deformations:
\begin{itemize}
\item[$\bullet$] \textit{Shear along the $x$-direction with the constant shear rate $\dot{\gamma}=4\times10^{10}$~s$^{-1}$}. For an elastic regime corresponding to small shear deformations~\cite{Landau_Lifshitz_1970}, the shear modulus $G$ is defined as the ratio of the shear stress $\tau_{xy}\equiv\sigma_{xy}$ to the shear strain $\gamma$ [see Figs.~\ref{fig_2}(a) and~\ref{fig_2}(d)]:
\begin{equation} \label{eq_G_modulus}
G =\frac{\tau_{xy}(t)}{\gamma(t)},
\end{equation}
where
\begin{equation} \label{eq_G_gamma}
\gamma(t) =\dot{\gamma}t.
\end{equation}
The shear stress simulations are realized via the method of non-equilibrium molecular dynamics in Lammps software package~\cite{Evans_Morriss_2008,Galimzyanov_Mokshin_2018}. In these simulations, atoms are located inside a triclinic simulation box. The shape of this simulation box changes due to the externally applied stress. For example, the simulation box changes the tilt in the $xy$-plane at shear applied along the $x$-direction. In Lammps, the tilt of the triclinic box is usually limited by half of the length of the box side parallel to $x$-axis. Therefore, the simulation box flips in the $xy$-plane at exceeding this limited length. These flips are cyclically repeated until the simulation completes. During each flip event, atoms are remapped into the flipped simulation box.
	
\item[$\bullet$] \textit{Uniaxial tension (along the $x$-axis) with the fixed tensile rate $4\times10^{10}$~s$^{-1}$}. The tensile modulus $E_{t}$ is defined as the ratio of the tensile stress $\sigma_{xx}$ to the tensile strain $\epsilon$ [see Figs.~\ref{fig_2}(a) and~\ref{fig_2}(b)]:
\begin{equation} \label{eq_E_modulus}
E_{t} =\frac{\sigma_{xx}(t)}{\epsilon(t)},
\end{equation}
where
\begin{equation} \label{eq_E_strain}
\epsilon(t) = \frac{L_{x}(t)-L_{x}^{(init)}}{L_{x}^{(init)}}.
\end{equation}
Here, $L_{x}^{(init)}$ is the length of the simulation box before the deformation, and $L_{x}(t)$ is this length at time $t$ after the start of deformation.

\item[$\bullet$] \textit{Uniaxial compression along the $x$-axis with the fixed rate $2\times10^{10}$~s$^{-1}$.} The compression modulus $E_{c}$ is defined by the ratio
\begin{equation} \label{eq_Ec_modulus}
E_{c} =\frac{\sigma_{xx}(t)}{\xi(t)},
\end{equation}
where $\sigma_{xx}$ is the compressive stress [see Figs.~\ref{fig_2}(a) and~\ref{fig_2}(c)], and $\xi$ is the compressive deformation:
\begin{equation} \label{eq_Ec_strain}
\xi(t) = \frac{L_{x}^{(init)}-L_{x}(t)}{L_{x}^{(init)}}.
\end{equation}
\end{itemize}

Note that the strain rates of the order $10^{10}$~s$^{-1}$ are achievable experimentally, for example, when a shock wave from an explosion impacts on a solid material~\cite{Bringa_2005,Shen_2012}. The strain with rate $\sim10^{10}$~s$^{-1}$ makes it possible to identify structural changes in the system at nanosecond time scales accessible to molecular dynamics simulations.

\section{Results and Discussion}

\subsection{Strain by shear}

\begin{figure*}[tbh]
	\centering
	\includegraphics[width=1.0\linewidth]{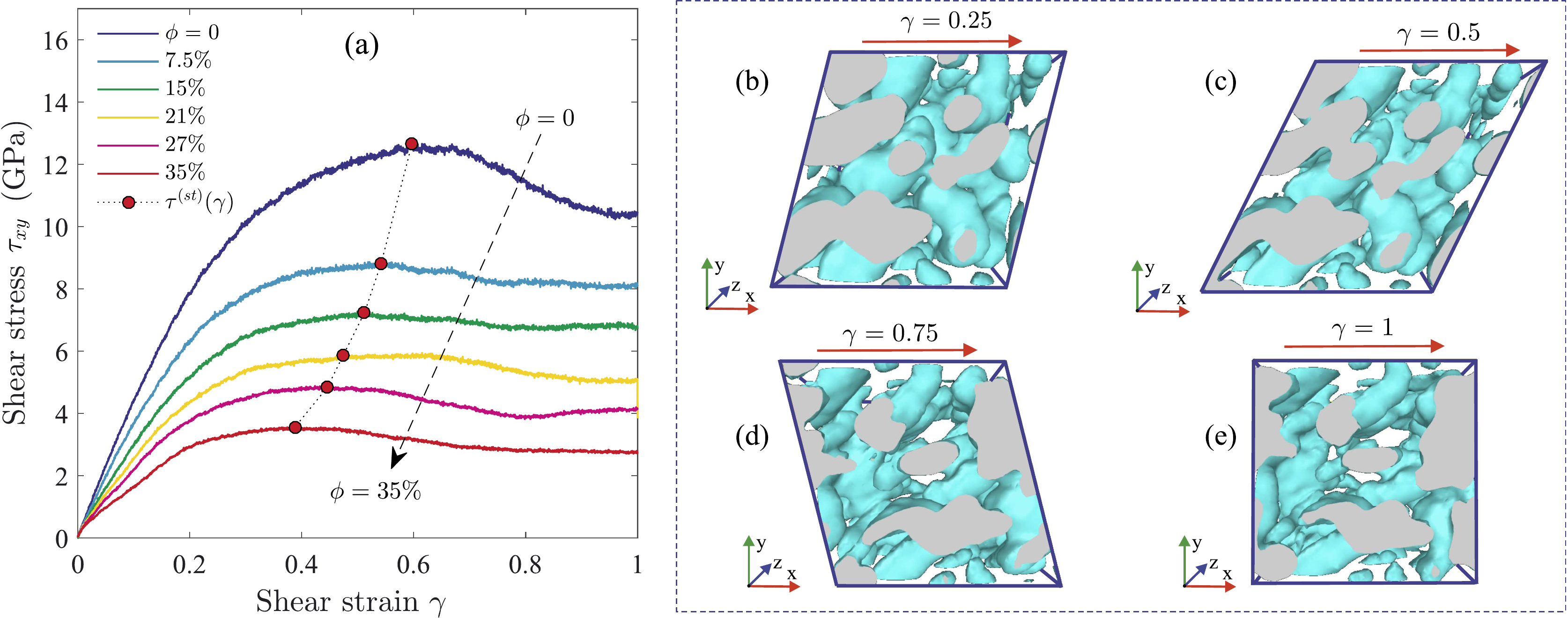}
	\caption{(color online) (a) Shear stress vs. shear strain curves for the homogeneous amorphous NiTi ($\phi=0$) and for the porous amorphous NiTi samples with various porosity $\phi$. Symbol (\textbullet) shows the peak shear strength $\tau^{(st)}$. Snapshots are given for the system with the porosity $35$\% at the different shear strains: (b) $\gamma=0.25$, (c) $\gamma=0.5$, (d) $\gamma=0.75$ and (e) $\gamma=1.0$. Pores are painted blue.}
	\label{fig_3}
\end{figure*}

Let us consider the samples of porous amorphous NiTi with different porosity $\phi$ and the sample of homogeneous amorphous NiTi ($\phi=0$) under shear deformation. The shear modulus $G$ for the considered samples is evaluated from the stress-strain curves, which are depicted in Fig.~\ref{fig_3}(a). As seen in Fig.~\ref{fig_3}(a), the stress-strain curves have a similar shape. All the curves have a pronounced maximum corresponded to yield strength and a plateau that characterizes the plastic deformation regime. For the homogeneous amorphous alloy, the shear modulus, yield strength, and residual shear strength in the plastic deformation regime take the values $G\simeq45$~GPa, $\tau^{(st)}\simeq12.7$~GPa and $\tau^{(res)}\simeq10.5$~GPa, respectively. For comparison, the found value of the shear modulus $G$ is approximately three times larger than the value $\approx18$~GPa experimentally estimated for the austenitic crystalline phase of NiTi at identical thermodynamic conditions~\cite{Sun_Li_2002}.
\begin{table}[tbh]
	\centering
	\caption{Mechanical characteristics of the system at shear deformation: shear modulus $G$, peak shear stress (yield strength) $\tau^{(st)}$, and residual shear strength $\tau^{(res)}$ in plastic deformation regime.}	
	\begin{tabular}{|c|c|c|c|}
		\hline
		$\phi$, \% & $G$, GPa & $\tau^{(st)}$, GPa & $\tau^{(res)}$, GPa \\
		\hline
		$0$ 	& $45\pm3.5$ & $12.7\pm1.1$ & $10.5\pm1.4$ \\
		$7.5$   & $37\pm3.2$ & $8.83\pm1.2$ & $8.13\pm1.1$ \\
		$15$ 	& $31\pm2.9$ & $7.23\pm1.1$ & $6.89\pm0.8$ \\
		$21$    & $26\pm2.4$ & $5.87\pm1.1$ & $5.13\pm0.7$ \\
		$27$    & $22\pm2.2$ & $4.85\pm0.9$ & $3.86\pm0.7$ \\
		$35$    & $19\pm2.1$ & $3.57\pm0.9$ & $2.79\pm0.5$ \\
		\hline
	\end{tabular}\label{tab_1}
\end{table}

The mechanical characteristics of the amorphous system strongly depend on the porosity of samples [see the values given in Table~\ref{tab_1}]. Namely, the values of the shear modulus, yield strength, and residual shear strength exhibit decrease with the increasing porosity $\phi$ in the plastic deformation regime. This tendency in the values of these mechanical characteristics is quite expected since the hardness of the system becomes less with increasing porosity. We find that the shear modulus for the system with the porosity $\phi=7.5$\% is approximately two times larger than in the case of the system with the porosity $\phi=35$\%. The reason for this is that the less porous samples are denser. In these samples, the pores have a shape close to spherical [see Fig.~\ref{fig_1}(c)]. The spherical pores are more resistant to external deformations than pores of any other shape~\cite{Smolin_Makarov_2016}. The samples with the high degrees of porosity, for example, with the porosity $\phi=35$\%, contain the percolating pore networks [see Fig. \ref{fig_1}(a)]. Figs.~\ref{fig_3}(b)-(e) show that shear-induced cavities of an elongated ellipsoid shape are formed inside the porous system. At large deformations, these cavities can coalesce and form nanoscale cracks. This effect is clearly visible in the video [see the video-file ``shear\-\_strain.avi'' in Supplementary Material], where the shear deformation of the porous system is visualized.

\subsection{Uniaxial tensile strain}

\begin{figure*}[tbh]
	\centering
	\includegraphics[width=1.0\linewidth]{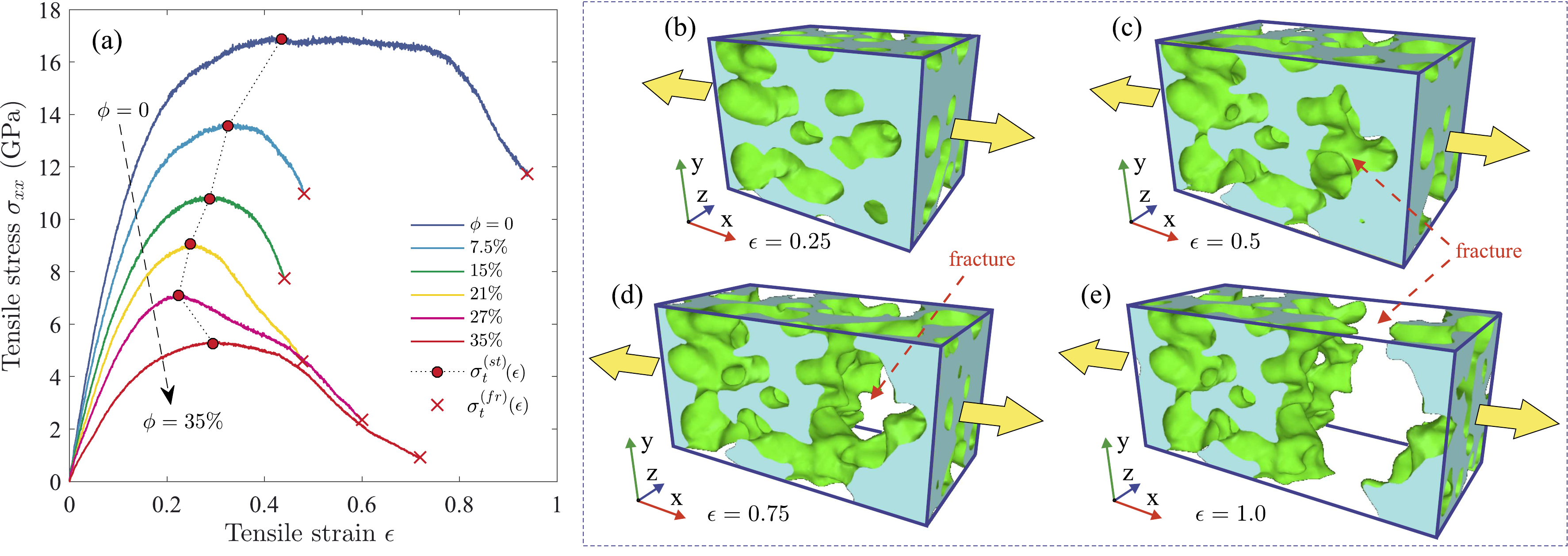}
	\caption{(color online) (a) Stress-strain diagram obtained at tensile of the homogeneous amorphous NiTi ($\phi=0$) and the porous amorphous NiTi with various porosity $\phi$. The sample with the porosity $35$\% at the different tensile strains: (b) $\epsilon=0.25$, (c) $\epsilon=0.5$, (d) $\epsilon=0.75$ and (e) $\epsilon=1.0$. The inner walls of the pores are colored green, whereas the dense regions without pores are in blue.}
	\label{fig_4}
\end{figure*}

Fig.~\ref{fig_4}(a) shows the stress-strain curves obtained by uniaxial tension of porous amorphous NiTi with different porosity $\phi$. The shape of these curves is typical for tensile stress. Namely, all the curves have the elastic region, the peak for ultimate tensile strength and the fracture point. The homogeneous amorphous sample exhibits the greatest elasticity and the resistance to the tensile deformation, as evidenced by the relatively large values of Young's modulus $E_{t}\simeq116$~GPa, the tensile strength $\sigma_{t}^{(st)}\simeq16.9$~GPa, and the fracture strain $\epsilon(\sigma_{t}^{(fr)})\simeq0.95$ [see Table~\ref{tab_2}]. It is important to note that the obtained value of the tensile modulus $E_{t}\simeq116$~GPa is in agreement with the experimental Young's modulus $\approx110$-$130$~GPa estimated for amorphous Ni and Ti-based alloys~\cite{Rouxel_Yokoyama_2015,Wang_2005}. On the other hand, the crystalline NiTi of the austenitic structure has the much lower Young's modulus $\approx41$-$75$~GPa at the temperature $300$~K~\cite{Nitinol_properties}.
\begin{figure}[tbh]
	\centering
	\includegraphics[width=0.55\linewidth]{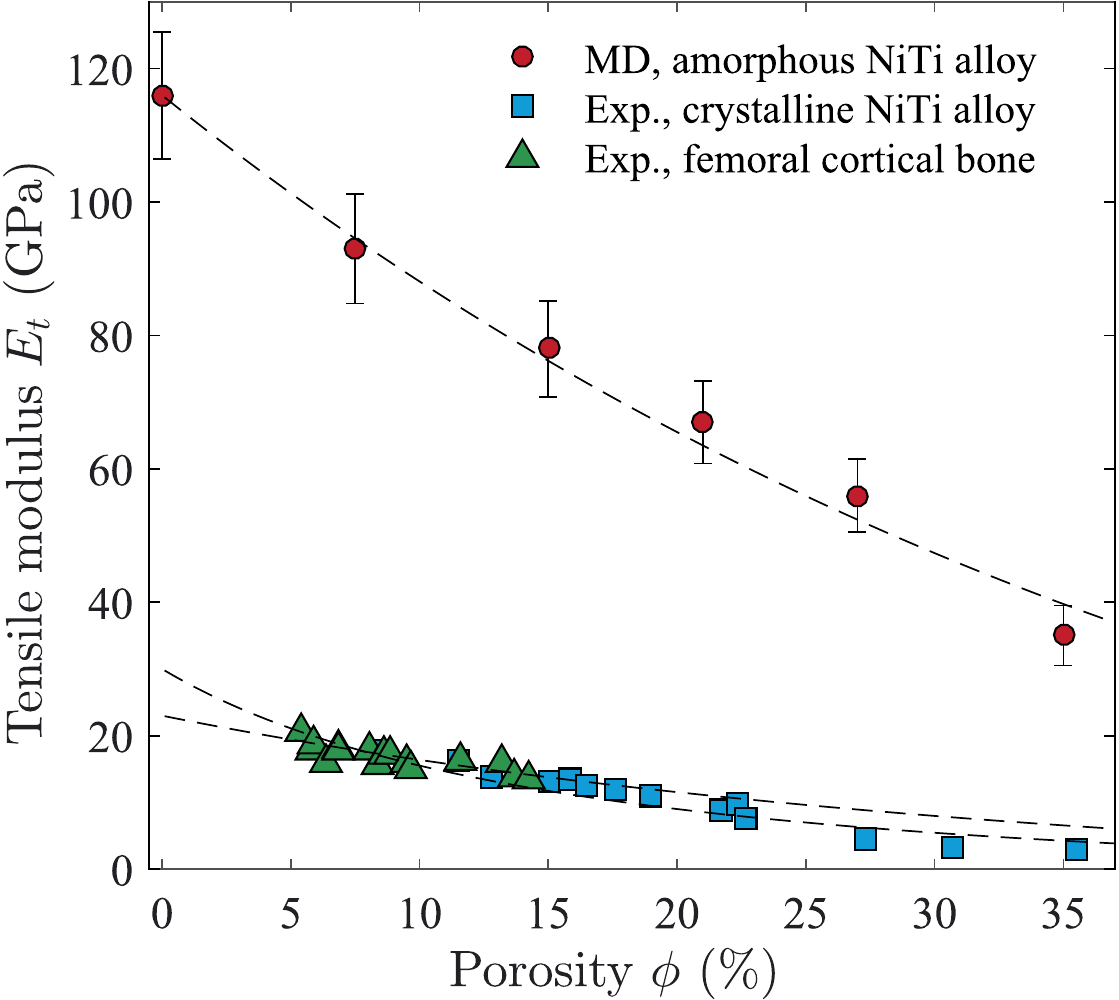}
	\caption{(color online) Young's modulus $E_{t}$ as function of the porosity $\phi$ estimated for the porous amorphous NiTi. The experimental data for the porous crystalline NiTi and the femoral cortical bone are measured at room temperature. Dashed curves are obtained by Eq.~(\ref{theory_N}).}
	\label{fig_5}
\end{figure}
\begin{table*}[tbh] 
	\centering
	\scriptsize
	\caption{Mechanical properties of the system at tensile and compressive deformations: tensile modulus $E_{t}$, ultimate tensile strength $\sigma_{t}^{(st)}$, strain $\epsilon^{(st)}$ at ultimate tensile strength, fracture tensile stress $\sigma_{t}^{(fr)}$ at the strain $\epsilon^{(fr)}$, compression modulus $E_{c}$, stress at the inception of the system densification $\sigma_{c}^{(dens)}$ at the strain $\xi^{(dens)}$.}
	\begin{tabular}{|c|c|c|c|c|c|c|c|c|}
		\hline
		$\phi$, \% & $E_{t}$, GPa & $\sigma_{t}^{(st)}$, GPa & $\epsilon^{(st)}$ & $\sigma_{t}^{(fr)}$, GPa & $\epsilon^{(fr)}$  & $E_{c}$, GPa & $\sigma_{c}^{(dens)}$, GPa & $\xi^{(dens)}$ \\
		\hline
		$0$   & $116\pm9.5$ & $16.9\pm2.1$ & $0.437$ & $11.5\pm1.2$  & $0.95$   & $125\pm7.4$ & $25.4\pm2.1$ & $0.322$ \\
		$7.5$ & $93\pm8.2$  & $13.6\pm1.5$ & $0.326$ & $10.6\pm1.2$  & $0.488$  & $100\pm6.6$ & $21.3\pm1.8$ & $0.352$ \\
		$15$  & $78\pm7.2$  & $10.8\pm1.2$ & $0.287$ & $7.55\pm0.8$  & $0.442$  & $83\pm5.2$  & $17.4\pm1.6$ & $0.383$ \\
		$21$  & $67\pm6.2$  & $9.06\pm1.0$ & $0.248$ & $4.14\pm0.4$  & $0.491$  & $71\pm5.0$  & $14.6\pm1.6$ & $0.393$ \\
		$27$  & $56\pm5.5$  & $7.11\pm0.8$ & $0.224$ & $2.35\pm0.2$  & $0.6$    & $63\pm4.5$  & $12.0\pm1.2$ & $0.392$ \\
		$35$  & $35\pm4.5$  & $5.27\pm0.6$ & $0.295$ & $0.91\pm0.15$ & $0.72$   & $46\pm3.8$  & $7.49\pm0.6$ & $0.369$ \\
		\hline
	\end{tabular}\label{tab_2}
\end{table*}

The mechanical characteristics of the system decrease significantly with increasing porosity [see Table~\ref{tab_2}]. So, with the increasing porosity from $7.5$\% to $35$\%, the Young's modulus $E_{t}$ and the tensile strength $\sigma_{t}^{(st)}$ decrease by $\approx2.5$ times, while the stress $\sigma_{t}^{(fr)}$ at the fracture point decreases by order of magnitude. These changes in the mechanical characteristics are due to decrease in the hardness of the system with increasing porosity. The complete destruction of the samples with the porosity $\phi=27$\% and $\phi=35$\% occurs at the deformations $\epsilon^{(fr)}\geq0.6$, while the samples with the porosity $\phi\leq21$\% are destroyed at the deformations $\epsilon^{(fr)}<0.5$. This is due to the fact that the pore network forms a relatively strong framework, which can stretch and, thereby, sustain large deformations without formation of nanoscale cracks. The cracks are usually for\-med at ultimate strain, and these cracks propagate in a direction perpendicular to the applied stress. This is clearly seen in Figs.~\ref{fig_4}(b)-(e), where fracture of the sample with the porosity $35$\% occurs through formation of single crack [see also the video-file ``tensile\_strain.avi'' in Supplementary Material]. Our results show that the system with the isolated pores has the smallest ultimate tensile strain $\epsilon^{(fr)}\simeq0.442$ at the porosity $15$\%. This is mainly due to absence of a formed percolating pore network.

In Fig.~\ref{fig_5}, the correspondence between the tensile modulus $E_{t}$ and the porosity coefficient $\phi$ calculated for porous amorphous NiTi is compared with the available experimental data obtained for porous crystalline NiTi~\cite{Elahinia_Hashemi_2012} and femoral cortical bone~\cite{Dong_Guo_2004}. The computed values of $E_{t}$ significantly exceeds the experimental Young's modulus in the porosity range $\phi\in[0;\,35]$\%. This is primarily due to the presence of an amorphous solid structure, which increases the strength and hardness of the alloy. Calculated $\phi$-dependence of the quantity $E_{t}$ holds the general tendency. Namely, the Young's modulus decreases with increasing porosity according to the Nielsen's equation~\cite{Nielsen_1984}:
\begin{equation}\label{theory_N}
E_{t}^{(N)}(\phi)=E_{t}^{(\phi=0)}\frac{\left(1-0.01\phi\right)^{2}}{1+0.01\phi(p-1)^{-1}}.
\end{equation}
In Eq.~(\ref{theory_N}), $p$ is the shape factor, which is usually taking as adjustable. Namely, this parameter is evaluated from fitting experimental or simulation data by Eq.~(\ref{theory_N}). According to Ref.~\cite{Nielsen_1984}, the shape factor takes small values when the pores form sharply edged networks, whereas its maximal value is $p=1$ for the case of isolated spherical pores. We find that the computed $\phi$-dependence of the Young's modulus $E_{t}$ is well reproduced by Eq.~(\ref{theory_N}) at the value $p=0.55$ [see Fig.~\ref{fig_5}]. This value corresponds to porous system with spherical pores. For comparison, the shape factor takes the value $p\simeq0.15$ for porous NiTi obtained by the crystalline powders sintering. The presence of defects and sharp edges in crystalline powders leads to formation of pores with non-trivial geometry, which ensures a small value of the shape factor $p$~\cite{Elahinia_Hashemi_2012}. Remarkable that the value of the shape factor $p$ found for the porous amorphous NiTi is close to the value $p\simeq0.42$ estimated for the femoral cortical bone, where the shape of pores is also close to spherical.
\begin{figure*}[tbh]
	\centering
	\includegraphics[width=1.0\linewidth]{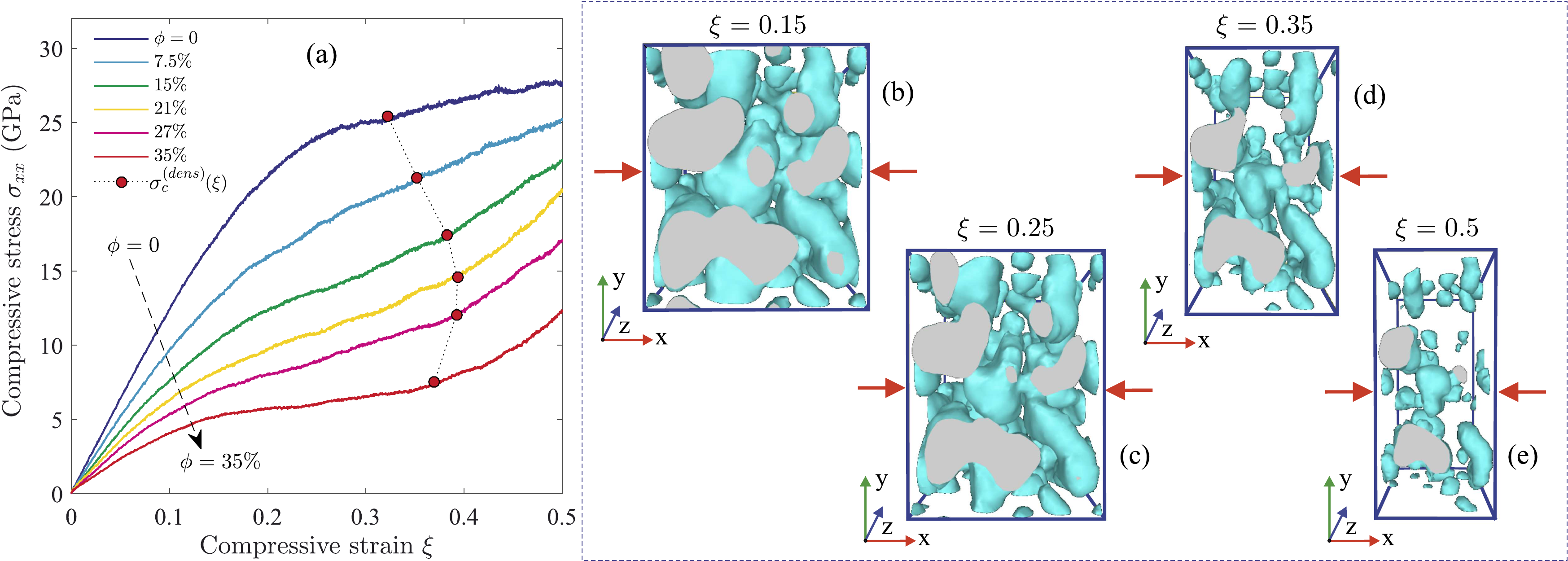}
	\caption{(color online) (a) Compression stress vs. strain curves calculated for the homogeneous amorphous NiTi ($\phi=0$) and for the porous amorphous NiTi with various porosity. Snapshots are shown the system with the porosity $35$\% at the different compressive strains: (b) $\xi=0.15$, (c) $\xi=0.25$, (d) $\xi=0.35$ and (e) $\xi=0.5$. Pores are painted blue.}
	\label{fig_6}
\end{figure*}

\subsection{Uniaxial compressive strain}

The samples of porous amorphous NiTi with various porosity were compressed to the strain $\xi=0.5$. Fig.~\ref{fig_6}(a) shows the set of the stress-strain curves obtained for the samples with various porosity. Each curve has a step, that is typical for materials with a porous structure, for example, metal foams and aerogels~\cite{Smith_Szyniszewski_2012,Ma_Zheng_2018}. This step arises due to collapse of the pores. As a result, a material becomes more dense. The points in the stress-strain curves associated with the transition to a more dense phase are marked in Fig.~\ref{fig_6}(a). As seen in Fig.~\ref{fig_6}(a), the transition to the densification region is smooth and weakly pronounced at the porosity $\phi\leq15$\%. This transition becomes more pronounced at the porosity $\phi>15$\% when a percolating pore network is formed inside the system. For example, Figs.~\ref{fig_6}(b)-(e) show the system with the porosity $\phi=35$\% at different strain. As seen from these figures and from the video-file ``compressive\_strain.avi'' in Supplementary Material, the percolating network of pores is collapsed and transformed into an ensemble of the isolated small-sized pores. As a result, the porous network is completely destroyed.
\begin{figure}[tbh]
	\centering
	\includegraphics[width=0.55\linewidth]{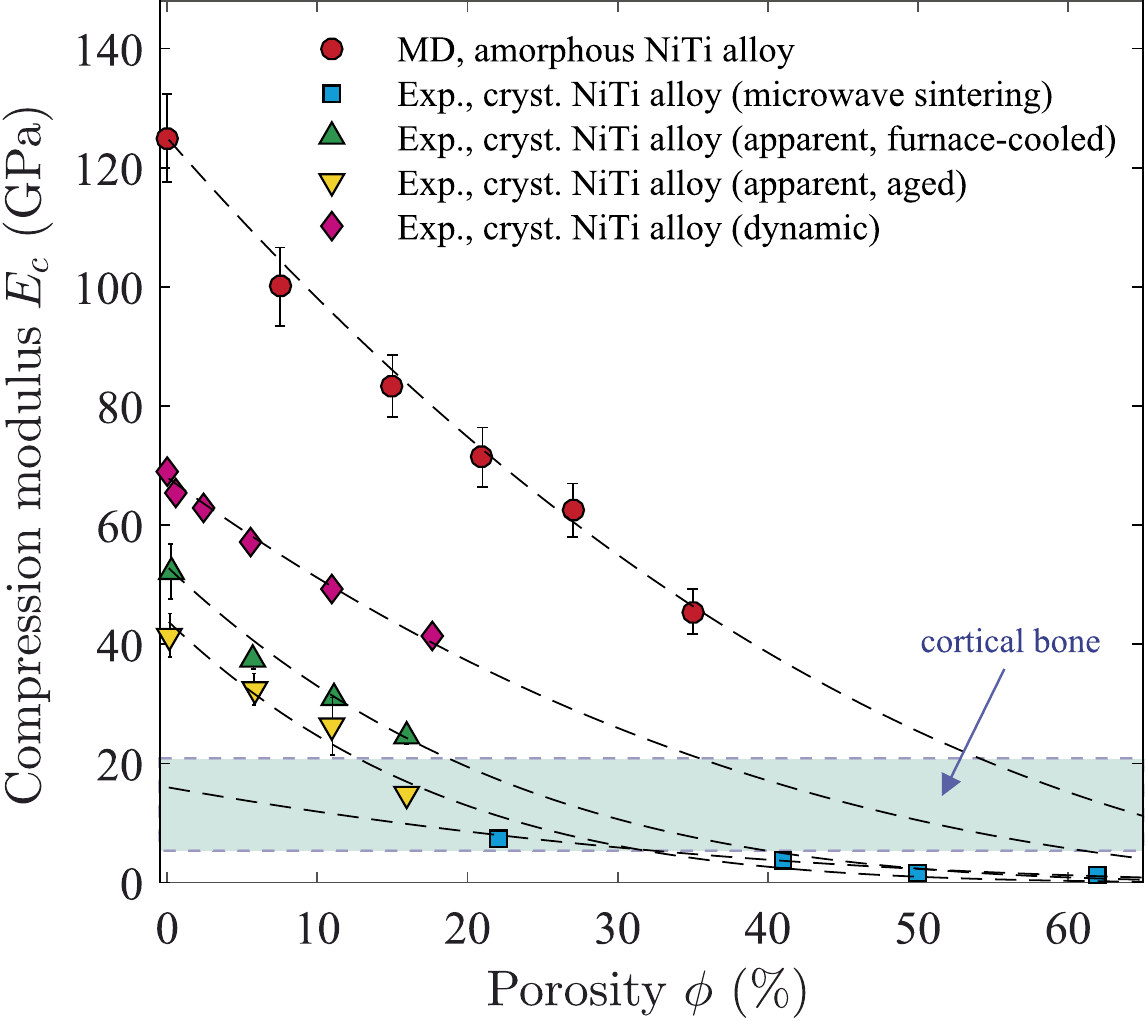}
	\caption{(color online) Compression modulus $E_{c}$ as function of the porosity $\phi$. Our results are compared with experimental data for porous crystalline NiTi prepared by microwave sintering~\cite{Xu_Bao_2015}, with the experimentally estimated dynamic and apparent Young's moduli at compression~\cite{Greiner_Oppenheimer_2005} as well as with the experimental data for cortical bone (region $5$--$21$~GPa)~\cite{Khan_Shyam_2014}. Dashed curves are the theoretical predictions by Eq.~(\ref{theory_B}).}
	\label{fig_7}
\end{figure}

The correspondence between the uniaxial compression modulus $E_{c}$ and the porosity $\phi$ is presented in Fig.~\ref{fig_7}. The values of the quantity $E_{c}$ are given in Table~\ref{tab_2}. The obtained values of the modulus $E_{c}$ are compared with the experimental data for porous crystalline NiTi synthesized at room temperature by various methods~\cite{Xu_Bao_2015,Greiner_Oppenheimer_2005}. Note that, for the considered porosity range, the value of the quantity $E_{c}$ estimated for the porous amorphous NiTi is higher in comparison to its crystalline analogue [see Fig.~\ref{fig_7}]. The $\phi$-dependence of the quantity $E_{c}$ is correctly reproduced by Bruno et al. equation~\cite{Smolin_Makarov_2016,Bruno_Efremov_2011}:
\begin{equation}\label{theory_B}
E_{c}^{(B)}(\phi)=E_{c}^{(\phi=0)}\left(1-0.01\phi\right)^{m}.
\end{equation}
Here, $m$ is the exponent to be determined. To evaluate the exponent $m$, experimental data or simulation results for $E_c(\phi)$ are approximated by Eq.~(\ref{theory_B}). The exponent $m$ characterizes the geometric features of pores at compression and the exponent can take values in the range from $2$ to $4$ ~\cite{Smolin_Makarov_2016}. For example, the value $m=2$ corresponds to a system with spherical pores, while the value $m=4$ characterizes a system with pores of pronounced non-spherical geometry. As seen in Fig.~\ref{fig_7}, we obtain a good agreement between the result of Eq.~(\ref{theory_B}) and the computed $\phi$-dependence of $E_{c}$ at the value $m=2.3$. For comparison, the experimental data for the crystalline analogue synthesized by powder sintering methods are well reproduced by Eq.~(\ref{theory_B}) at $m\simeq3.8\pm0.4$.

It is noteworthy that the compression modulus $E_{c}$ is higher than the tensile modulus $E_{t}$ at the same porosity. According to Table~\ref{tab_2}, the difference between the values of the moduli $E_t$ and $E_c$ decreases slightly when the porosity $\phi$ increases from $0$ to $21$\%. Further, this difference increase with the increasing porosity from $21$\% to $35$\% mainly due to the predominance of $E_c$. We assume that this non-monotonic behavior of the $\phi$-dependences of $E_t$ and $E_c$ is caused by coalescence of the isolated pores and formation of the percolating network of pores at the porosity $\phi>21$\% [see Fig.~\ref{fig_1}]. Namely, the atoms form a steady framework with open-porous channels within the amorphous system. This framework creates a weak internal mechanical stress in the system that counteracts to destruction of the porous structure during compression.

\section{Conclusions}

In the present work, the mechanical properties of the mesoporous amorphous NiTi alloy were studied by the method of non-equilibrium molecular dynamics simulations. Samples of the system with different porosity are analysed at uniaxial tension, uniaxial compression, and uniform shear. The stress-strain curves are found, and the values of the main mechanical characteristics as functions of system porosity are calculated. It was shown that the Young's modulus computed at tension and compression of porous amorphous NiTi exceeds the experimental values for crystalline analogues with the same porosity. This indicates that the resistance of the porous amorphous system to external deformations is higher than for the case of the porous crystalline system. The Young's modulus tends to decrease with increasing porosity according to Nielsen's power-law (in the case of tensile stress) and according to power-law of Bruno et al. (in the case of compressive stress). This tendency is in good agreement with the results of the available experimental data obtained for the porous crystalline NiTi and the femoral cortical bone.

We find that the degree of the system porosity is correlated with the pore geometry. The amorphous alloy with the porosity $7.5$\%$\leq\phi\leq15$\% contains mainly isolated pores of the spherical shape. At the porosity $\phi>15$\%, the pores form a percolating network. It has been established that the amorphous solid NiTi with the isolated spherical pores better resists to compression, while it is less resistant to tensile and shear deformations. The amorphous NiTi alloy with a percolating network of pores is able to withstand large deformations at tension and shear. The results of the present work complement existing studies of the mechanical properties of Ni and Ti-based alloys with the mesoporous structure, and these results can be demanded in the development of design methods of porous amorphous alloys with the necessary characteristics of the porosity.

\section*{Acknowledgement}
This study is supported by the Russian Science Foundation (project No. 19-12-00022); the computational part of the study is supported by the Russian Foundation for Basic Research (project no. 18-02-00407-a). AVM acknowledges the Foundation for the Advancement of Theoretical Physics and Mathematics ``BASIS'' for support.

\end{document}